\documentclass[sigconf]{acmart}

\usepackage{enumitem} 

\AtBeginDocument{%
  }

\copyrightyear{2025}
\acmYear{2025}
\setcopyright{rightsretained}
\acmConference[CHI EA '25]{Extended Abstracts of the CHI Conference on Human Factors in Computing Systems}{April 26-May 1, 2025}{Yokohama, Japan}
\acmBooktitle{Extended Abstracts of the CHI Conference on Human Factors in Computing Systems (CHI EA '25), April 26-May 1, 2025, Yokohama, Japan}\acmDOI{10.1145/3706599.3720174}
\acmISBN{979-8-4007-1395-8/2025/04}

\begin{document}




\title{Enhancing Autonomous Vehicle-Pedestrian Interaction in Shared Spaces: The Impact of Intended Path-Projection}


\author{Le Yue}
\email{lyue0331@uni.sydney.edu.au}
\orcid{0009-0006-3976-7120}
\affiliation{Design Lab, Sydney School of Architecture, Design and Planning,
  \institution{The University of Sydney}
  \city{Sydney}
  \state{NSW}
  \country{Australia}
}

\author{Tram Thi Minh Tran}
\email{tram.tran@sydney.edu.au}
\orcid{0000-0002-4958-2465}
\affiliation{Design Lab, Sydney School of Architecture, Design and Planning,
  \institution{The University of Sydney}
  \city{Sydney}
  \state{NSW}
  \country{Australia}
}
\author{Xinyan Yu}
\email{xinyan.yu@sydney.edu.au}
\orcid{0000-0001-8299-3381}
\affiliation{Design Lab, Sydney School of Architecture, Design and Planning
  \institution{The University of Sydney}
  \city{Sydney}
  \state{NSW}
  \country{Australia}
}

\author{Marius Hoggenmueller}
\email{marius.hoggenmueller@sydney.edu.au}
\orcid{0000-0002-8893-5729}
\affiliation{Design Lab, Sydney School of Architecture, Design and Planning
  \institution{The University of Sydney} 
  \city{Sydney}
  \state{NSW}
  \country{Australia}
}

\renewcommand{\shortauthors}{Yue et al.}


\begin{abstract} 

External Human-Machine Interfaces (eHMIs) are critical for seamless interactions between autonomous vehicles (AVs) and pedestrians in shared spaces. However, they often struggle to adapt to these environments, where pedestrian movement is fluid and right-of-way is ambiguous. To address these challenges, we propose PaveFlow, an eHMI that projects the AV's intended path onto the ground in real time, providing continuous spatial information rather than a binary stop/go signal. Through a VR study (N=18), we evaluated PaveFlow's effectiveness under two AV density conditions (single vs. multiple AVs) and a baseline condition without PaveFlow. The results showed that PaveFlow significantly improved pedestrian perception of safety, trust, and user experience while reducing cognitive workload. This performance remained consistent across both single and multiple AV conditions, despite persistent tensions in priority negotiation. These findings suggest that path projection enhances eHMI transparency by offering richer movement cues, which may better support AV-pedestrian interaction in shared spaces.

\end{abstract}


\begin{teaserfigure}
  \centering
  \includegraphics[width=\linewidth]{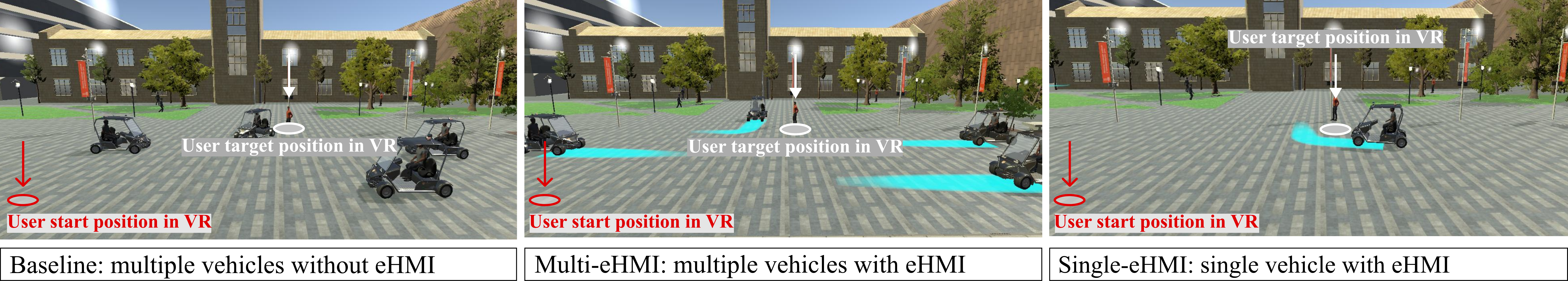}
\caption{The three experimental conditions in this study: (Left) \textbf{Baseline} – multiple vehicles without eHMI, (Middle) \textbf{Multi-eHMI} – multiple vehicles with eHMI, (Right) \textbf{Single-eHMI} – single vehicle with eHMI.}
  \label{fig:teaser}
  \Description{Interior virtual reality scene depicting user start and target positions across three experimental conditions in a shared space environment. From left to right: Baseline, featuring multiple vehicles without eHMI; Multi-eHMI, with multiple vehicles equipped with eHMI; and Single-eHMI, showing a single vehicle with eHMI. The positions are clearly marked from the user start positions to the target positions, demonstrating the setup for each condition.}
\end{teaserfigure}

\maketitle

\section{Introduction}

Shared spaces eliminate traditional vehicle-pedestrian boundaries, necessitating real-time right-of-way negotiation~\cite{ruiz2017shared, JENS2021102592, karndacharuk2014review, predhumeau2021pedestrian, kabtoul2021proactive}, with no party holding absolute priority~\cite{9552508}. This process, known as co-navigation, involves continuous mutual adjustment between road users~\cite{yu2023your}. In such environments, clear communication becomes essential for safe and efficient interactions. 
For autonomous vehicles (AVs), External Human-Machine Interfaces (eHMIs) have been proposed to clarify AV intentions and improve AV-pedestrian interactions~\cite{dey2020taming}. 
However, most existing eHMIs are designed for structured roads~\cite{wang2022pedestrian}, typically focusing on one-to-one communication rather than the dynamic, multi-user co-navigation characteristic of shared spaces~\cite{predhumeau2021pedestrian, wang2022pedestrian}.

In shared spaces, AVs need to dynamically respond to pedestrians in real time~\cite{10.1007/978-3-030-28619-4_25}. Current AVs rely on internal path adjustment algorithms to achieve this, but the subtlety of their motion cues in low-speed shared spaces can render these adjustments opaque to pedestrians.
Furthermore, explicit communication via eHMIs is often reduced to binary responses—either yielding or not yielding when paths may cross~\cite{zhang2024external}—without accommodating more nuanced interactions. Moreover, AV communication typically does not adapt to the continuous nature of pedestrian movements (e.g., altering paths), which limits real-time negotiation in shared environments~\cite{wang2022pedestrian}.


To address this gap, we introduce PaveFlow, an eHMI that externalises an AV's intended path and responds to pedestrians in real-time. By making an AV's future movements more transparent, PaveFlow aims to improve AV-pedestrian interactions in shared spaces. 
To evaluate the effectiveness and scalability of PaveFlow, we conducted a within-subjects Virtual Reality (VR) study (N=18) with three experimental conditions: PaveFlow visualisation in two AV density situations (single vs. multiple AVs) and a baseline condition without PaveFlow. We analysed participant feedback on perceived safety, trust, workload, and user experience to assess PaveFlow's impact on co-navigation. Results showed that PaveFlow significantly enhanced pedestrian confidence and trust, lowered cognitive workload, and performed consistently across single and multi-AV conditions. However, in multi-AV settings, participants reported confusion as different AV motion states appeared simultaneously in PaveFlow projections.

The contributions of this paper are twofold:
\begin{itemize}[topsep=0pt]
    \item Introduces PaveFlow as a design concept for a scalable, adaptive eHMI in shared spaces.
    \item Provides empirical insights into how externalised path information shapes AV-pedestrian interactions, informing future eHMI development.
\end{itemize}

\section{Related Work}

\subsection{AV-Pedestrian Interaction in Shared Spaces}

Prior research on AV-pedestrian interaction has mainly centred on algorithmic approaches to pedestrian handling. Approaches such as game-theoretic models~\cite{camara2021evaluating}, predictive models~\cite{shoman2024transformer}, and proxemics-based methods~\cite{camara2021space} are theoretically effective but underperform in practice~\cite{camara2022unfreezing}. AV motion planning algorithms often fail to handle pedestrian unpredictability and lack effective communication, leading to overly cautious behaviours like the \textit{`freezing robot problem'}~\cite{crosato2021human}. 

While evidence shows that eHMIs improve AV-pedestrian interactions~\cite{rouchitsas2019external}, 
few studies focus on eHMIs designed specifically for shared spaces~\cite{wang2022pedestrian, Hoggenmueller2022frontiers}. Existing eHMIs are often developed for structured environments with clear right-of-way rules and one-to-one interactions, making them ill-suited for shared spaces where priority is ambiguous and multiple road users must coordinate movement~\cite{predhumeau2021pedestrian, wang2022pedestrian}. This lack of clear communication increases uncertainties and lowers efficiency~\cite{jayaraman2019pedestrian,she2021shaping,verma2019pedestrians}.

Effective co-navigation in shared spaces demands eHMIs that facilitate complex, multi-user negotiations~\cite{yu2024understanding, singamaneni2024survey}. 
Existing eHMI designs primarily rely on communication signals such as LED light bars~\cite{habibovic2018communicating, lee2022learning, de2019external}, with 97\% of them focusing on conveying the vehicle's yielding intent~\cite{dey2020taming}. While this information is crucial for structured roads and crossing scenarios, it is less effective in shared spaces, where seamless co-navigation depends on pedestrians being able to anticipate AV movements more granularly and adjust their own accordingly. A notable attempt to improve AV-pedestrian communication beyond yielding intent is \textit{Gazing Eyes}~\cite{gui2022gazing}, which uses anthropomorphic robotic eyes to indicate the vehicle's fine-grained moving direction. However, eye-based communication can be subtle and difficult to notice. 

\subsection{Projection-Based eHMI}

\sloppy Existing projection-based eHMIs convey AV intentions through ground-projected symbols like stop lines~\cite{mercedes2015, nguyen2019projection, tran2024exploring}, symbols~\cite{shah2022study}, and directional arrows~\cite{dey2020taming, mitsubishi2015indicators}. While factors such as adverse weather~\cite{rettenmaier2019passing}, bright daylight~\cite{nguyen2019projection}, glare, and uneven surfaces~\cite{rettenmaier2019passing} can affect visibility, projection-based eHMIs remain highly adaptable and practical in shared spaces. \citet{nguyen2019projection} highlighted several advantages: they are visible to multiple pedestrians at once, potentially from different viewing angles, and they utilise the road surface, a familiar medium for conveying traffic information. Building on these strengths, our study explores a dynamic path projection concept that enhances AV-pedestrian interaction.

\section{METHODOLOGY}


\subsection{Design Concept}
\begin{figure*}
  \centering
  \includegraphics[width=0.9\linewidth]{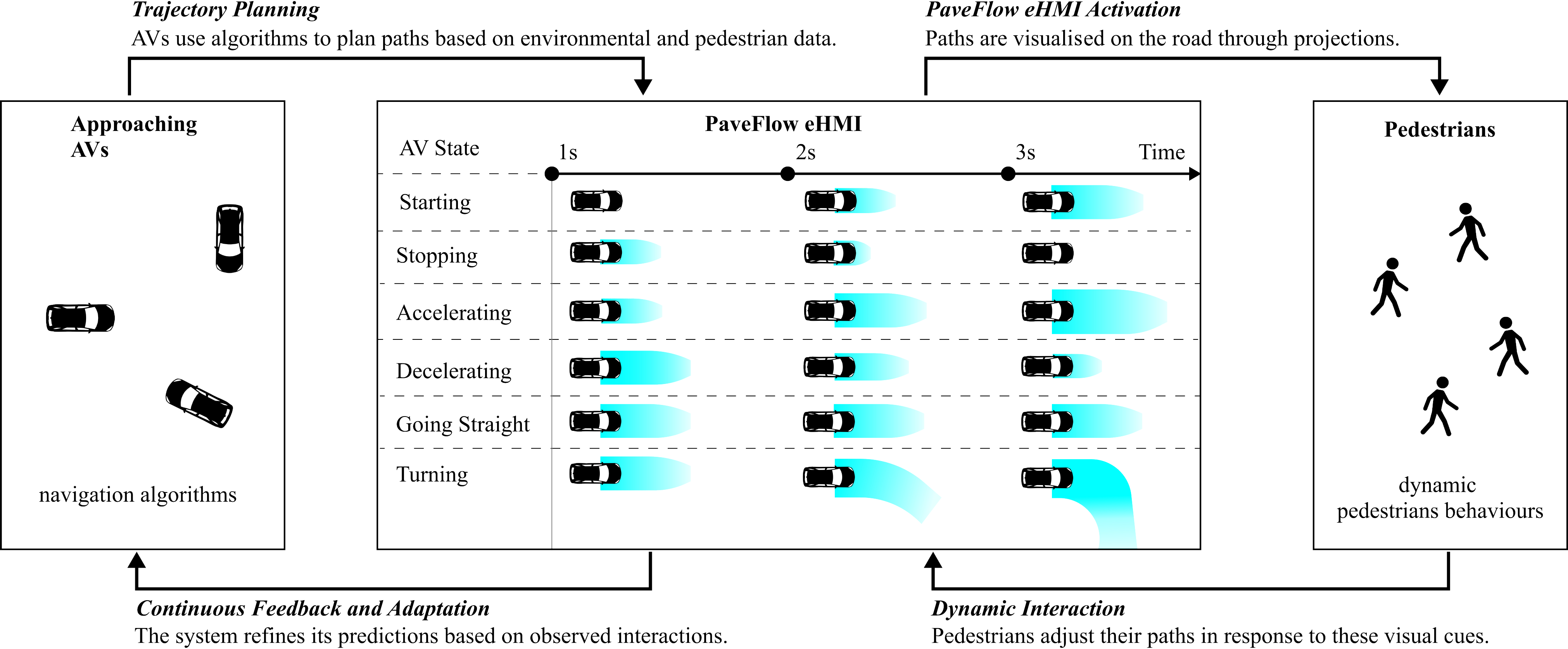}
  \caption{Overview of design concept which integrates AV navigation algorithms with a path projection eHMI to visualise intended trajectories in real time. As pedestrians adjust their movement in response, the concept enables dynamic right-of-way negotiation.}
  \label{fig:design-concept}
  \Description{system flowchart illustrates the operational architecture, featuring components 'Approaching AVs', 'PaveFlow eHMI', and 'Pedestrians'. In this system, AVs utilize algorithms to plan paths based on environmental and pedestrian data, which are fed into the 'PaveFlow eHMI'. The paths are visualised on the road through projections, communicated to 'Pedestrians', who adjust their paths in response to these visual cues. These adjustments are captured by the 'PaveFlow eHMI', allowing the system to refine its predictions based on observed interactions, which are subsequently received by 'Approaching AVs'. The diagram also depicts the visualisation of AV motion states within 'PaveFlow', demonstrating how it changes over time.}
\end{figure*}

The design concept, \textit{PaveFlow}, integrates AV's 
navigation algorithms~\cite{HUANG2024127763,yu2020risk} with a path projection eHMI,
as shown in \autoref{fig:design-concept}. It dynamically projects the AV's intended trajectories onto the ground and adjusts them in real time based on pedestrian movements. PaveFlow was designed around three principles derived from a detailed analysis of co-navigation challenges in shared spaces~\cite{wang2022pedestrian}.


\begin{itemize}
\item \textit{Continuity}: Keep AV intentions constantly updated, helping pedestrians and AVs adjust in real time.
\item \textit{Predictability}: Offer clear visual cues that help pedestrians easily understand AV intentions.

\item\textit{Flexible Right-of-Way}:
Allow AVs and pedestrians to flexibly share the right of way. Instead of AVs always yielding to pedestrians, PaveFlow enables a more balanced interaction where pedestrians can decide to yield based on the AV’s projected path.

\end{itemize}




PaveFlow was developed through an iterative process, beginning with low-fidelity sketches and evolving into a high-fidelity 3D Unity prototype where an AV navigation algorithm was developed. 
Based on this algorithm, the system projects the AV's expected path up to three seconds ahead (see \autoref{fig:design-concept}, middle), providing pedestrians with clear, anticipatory visual cues. 

We selected cyan to convey the AV’s intended trajectory, as research supports it as optimal colour for eHMI: it is distinguishable~\cite{werner2018new}, neutral for yielding intent~\cite{dey2020color}, and intuitive for pedestrian crossings~\cite{bazilinskyy2020external}.

\subsection{Experimental Design and Hypotheses}

The experiment employed a within-subjects design in which each participant experienced three conditions (see \autoref{fig:teaser}): 
\begin{itemize}
    \item \textit{Baseline}: Multiple AVs and no eHMI
    \item \textit{Multi-eHMI}: Multiple AVs with eHMI
    \item \textit{Single-eHMI}: Single AV with eHMI
\end{itemize}
The \textit{Baseline} condition serves as a reference for AV-pedestrian interactions in realistic shared spaces with multiple vehicles, ensuring findings remain applicable to real-world scenarios. Using a multi-AV baseline allows us to assess whether eHMI remains effective when pedestrians must interpret multiple AVs simultaneously, rather than in isolated AV-pedestrian encounters. Since AV behaviour was fully controlled in VR, this design choice does not introduce additional variability. The \textit{Multi-eHMI} condition evaluates eHMI effectiveness in these high-density settings. The \textit{Single-eHMI} condition was added to examine how eHMI functions in simpler AV-pedestrian interactions.

Based on the theory of shared mental models~\cite{scali2019shared}, which suggests that transparent and predictable communication helps individuals anticipate system behaviour, reduce uncertainty, and build trust, we propose the following hypotheses:

\begin{itemize}
\item {\textit{H1 (Effectiveness)}}: PaveFlow improves AV-pedestrian interactions by enhancing safety and trust, reducing cognitive workload, and improving user experience compared to the baseline condition without eHMIs.
\item {\textit{H2 (Scalability)}}: PaveFlow maintains its effectiveness in both single and multiple AV conditions.
\end{itemize}





\subsection{Apparatus}

PaveFlow was evaluated through a VR-based experiment. The virtual environment was modelled to replicate an actual shared space within our university campus. The AV used in the experiment was a small, motorised vehicle, typically used in shared spaces~\cite{wang2022pedestrian, florentine2016pedestrian}. The AV operated at low speeds, up to 10 km/h. Originally, the AV was designed to predict pedestrian movement paths and adjust its behaviour dynamically. However, in the experiment, we simplified this logic: the AV only adjusted its speed and projected its path based on its distance to pedestrians, without predicting their future movement. This simplification ensured consistent and repeatable interactions, avoiding variability from prediction errors.
The study took place in an open floor area with an 11×13-metre movement zone. A Meta Quest 2 VR headset was used to create an immersive simulation.

\subsection{Data Collection and Analysis}


\subsubsection{Quantitative Measures:} 
\begin{itemize} \item {\textit{Perceived Safety}}: Assessed using a 5-point Likert Scale adapted from prior research~\cite{hollander2022take}.
\item {\textit{Trust in Automation}}: Measured using the Trust in Automation Scale~\cite{jian2000foundations}.
\item {\textit{Workload}}: Evaluated using the NASA Task Load Index~\cite{cao2009nasa}.
\item {\textit{User Experience}}: Measured using the short version of the User Experience Questionnaire~\cite{schrepp2017design}.

\end{itemize}

Quantitative data were analysed using descriptive statistics to summarise trends and inferential statistics to test for differences across conditions. A global test across all three conditions was not conducted because \textit{Baseline} and \textit{Single-eHMI} differ in both the number of AVs and the presence of eHMI, making direct comparisons less meaningful. Instead, two separate tests were conducted: one-way repeated measures ANOVA for normally distributed data (\textit{Baseline} and \textit{Multi-eHMI}) and the Wilcoxon signed-rank test for non-parametric data (between \textit{Multi-eHMI} and \textit{Single-eHMI}). A significance threshold of p < 0.05 was applied for all tests.

\subsubsection{Qualitative Measures:} 

Qualitative data were collected through semi-structured interviews conducted post-condition (4-8 minutes each) and post-study (20-30 minutes) to assess participants' perceptions of the PaveFlow eHMI, user experience, and AV behaviour. The interviews focused on participant's overall experience of interacting with the AVs, the different conditions, and suggestions for design improvement. Thematic analysis was employed to extract key themes and insights, helping to identify recurring patterns and refine the understanding of participant feedback.





\subsection{Participants}
Eighteen participants (aged 18–34) were recruited using posters, flyers, and social media. The group comprised 16 women, 1 men, and 1 person who did not disclose their gender. Regarding VR and AV familiarity, 8 participants had no prior VR experience and 10 participants were somewhat familiar with AVs~(e.g., through popular media).


\subsection{Study Procedure}

The experimental procedure comprised four phases to ensure participant familiarity and minimise learning effects. In the familiarisation phase, participants adapted to the VR environment and navigation tasks without moving vehicles. During the practice round, they navigated to a stationary target without distractions to learn the VR controls. In the actual experiment, participants completed three randomised conditions, each lasting about three minutes, followed by a questionnaire and brief feedback during a short interview. Finally, the post-experiment session involved a semi-structured interview to discuss experiences across conditions and gather suggestions for improvement. This study was carried out following the ethical approval granted by the University of Sydney (project protocol 2023/HE000434).



\section{RESULTS}

\begin{figure*}
  \centering
  \includegraphics[width=0.8\linewidth]{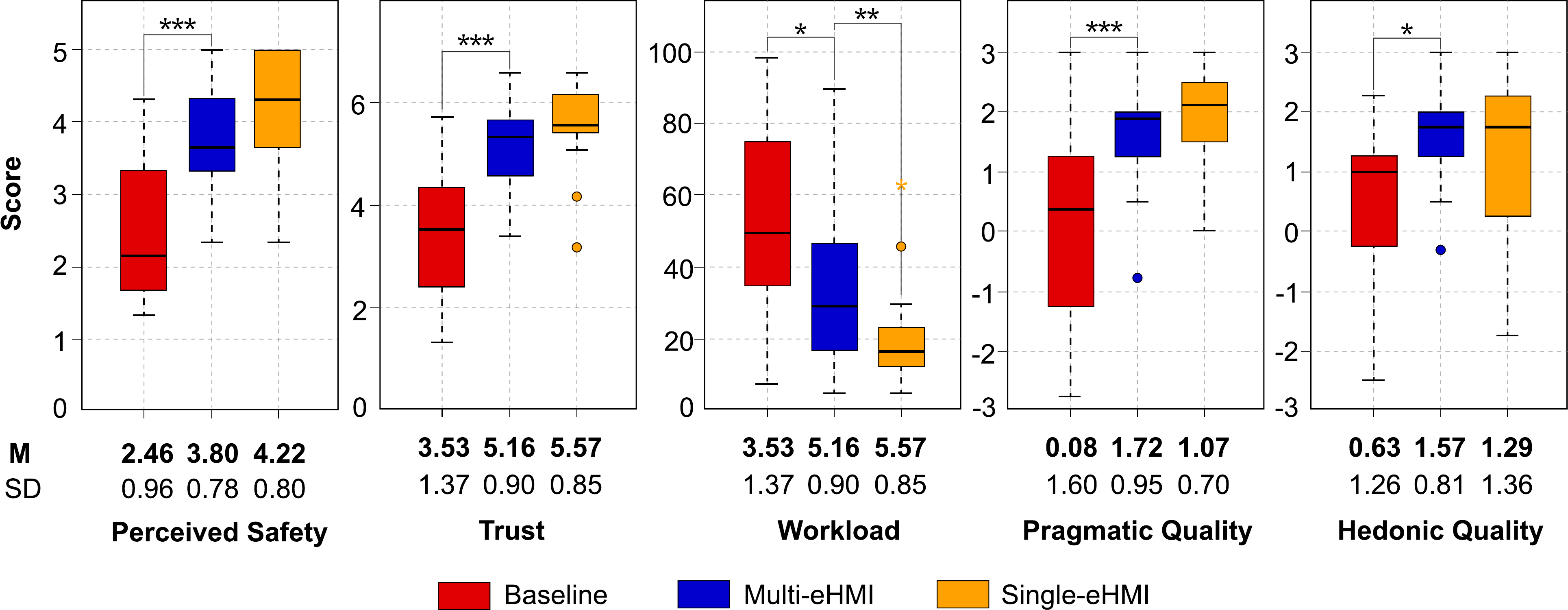}
  \caption{Boxplots show Descriptive Means (M) and Standard Deviations (SD) of Perceived Safety, Trust, Workload, Pragmatic Quality, and Hedonic Quality across conditions. Statistical significance was tested using ANOVA (Baseline vs. Multi-eHMI) and Wilcoxon (Multi-eHMI vs. Single-eHMI). *: p $\leq$ .05, **: p $\leq$ .01, ***: p $\leq$ .001}
  \label{fig:overview_chart}
  \Description{Boxplots of user evaluations in three experimental conditions—Baseline (red), Multi-eHMI (blue), and Single-eHMI (yellow)—for five different metrics: Perceived Safety, Trust, Workload, Pragmatic Quality, and Hedonic Quality. Mean values and standard deviations for each condition are listed below the respective boxplots. Perceived Safety and Trust: Exhibit increased scores in more complex eHMI setups. Workload: Highlights significant variance in the Baseline condition, suggesting diverse user experiences. Pragmatic and Hedonic Quality: Show variances in user satisfaction, with the Single-eHMI condition generally scoring higher on hedonic measures. Statistical significance, denoted by asterisks, indicates the strength of differences between conditions. 
}
\end{figure*}

\autoref{fig:overview_chart} presents the descriptive statistics along with statistical significance indicators for comparisons across conditions.

\subsection{Perceived Safety}


A one-way repeated measures ANOVA comparing \textit{Baseline} and \textit{Multi-eHMI} revealed a significant effect of eHMI presence (F(1,18) = 18.133, p < 0.001), with \textit{Multi-eHMI} leading to higher Perceived Safety. A Wilcoxon signed-rank test comparing \textit{Multi-eHMI} and \textit{Single-eHMI} showed no significant difference (Z = 1.570, p = 0.116).

Qualitative feedback: In \textit{Baseline}, participants felt `surrounded' (P13) and `confused' (P10, P15) from the lack of cues. Many expressed uncertainty about whether the vehicle would yield (n=8) and hesitated to proceed walking due to fear of collision (n=5). 
In \textit{Multi-eHMI}, pedestrians showed improved awareness of AV actions like `turning' and `going straight' (n=8) compared to \textit{Baseline}. However, they struggled to accurately identify changes like `stopping' or `slowing down' when multiple eHMIs were present (n=7). To improve comprehension and enhance pedestrian responsiveness, participants suggested to incorporate auditory signals for `stopping' (n=6) and providing detailed speed information for `accelerating/decelerating' (n=8).
In \textit{Single-eHMI}, participants actively adjusted their paths and speed (n=10) in response to changes in the AV’s movement. Comments such as `I believe the vehicle won’t hit me’ (P18) suggest that the PaveFlow’s visual cues provided reassurance and helped pedestrians navigate safely.

\subsection{Trust in Automation}
A one-way repeated measures ANOVA comparing \textit{Baseline} and \textit{Multi-eHMI}, revealed a significant effect of eHMI presence (F(1,18) = 17.713, p < 0.001), with \textit{Multi-eHMI} leading to higher trust. A Wilcoxon signed-rank test comparing \textit{Multi-eHMI} and \textit{Single-eHMI} showed no significant difference (Z = 1.659, p = 0.097).


Qualitative feedback: In \textit{Baseline}, changes in AV motion states (e.g., yielding or turning) caused pedestrian anxiety (n=3). 
For instance, pedestrians often misinterpreted `yielding to stop' as an `AV malfunction' (P18), leading to mistrust and tense behaviour. 
Reactions varied: some waited for all AVs to pass, while others hurried to avoid collisions. 
In \textit{Multi-eHMI}, 
pedestrian trust generally increased.
However, it is important to note that within this environment, 
trust decreased during AV state transitions due to challenges in quickly interpreting PaveFlow information (n=9).
Participants (e.g., P12, P14, P18) suggested that familiarity with PaveFlow could enhance usage confidence.
In \textit{Single-eHMI}, pedestrians navigated at a normal pace, instinctively interpreting PaveFlow as the AV's path (n=14). 

\subsection{Workload}
\begin{figure*}
  \includegraphics[width=0.75\linewidth]{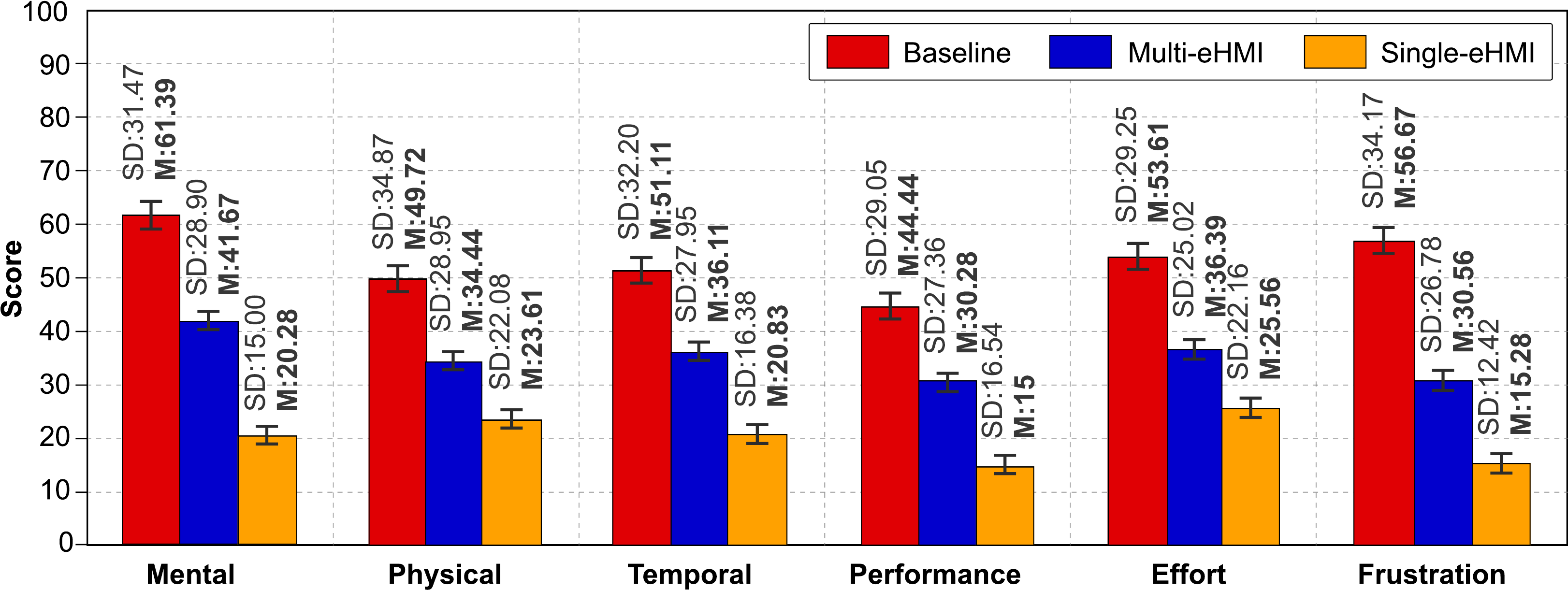}
  \caption{Descriptive statistics of NASA-TLX workload dimensions across the three study conditions.}
  \label{fig:workload}
  \Description{ Bar chart displays user scores on the Y-axis from 0 to 100 for six metrics: Mental, Physical, Temporal, Performance, Effort, and Frustration. The X-axis lists these metrics. Scores are shown in three color-coded conditions: Baseline (red), Multi-eHMI (blue), and Single-eHMI (yellow). Each bar includes the mean score (M) and standard deviation (SD) values. Mental scores show that the Single-eHMI condition consistently has the highest scores, peaking near 90. Physical and Temporal scores are relatively close across all conditions, with Single-eHMI marginally higher. Performance scores slightly increase in the Single-eHMI condition. Effort shows a notable decrease in the Single-eHMI condition compared to others. Frustration levels are consistently lower in the Single-eHMI condition}
\end{figure*}


\autoref{fig:workload} presents descriptive workload ratings across conditions for each NASA-TLX dimension.
For overall workload, a one-way repeated measures ANOVA comparing \textit{Baseline} and \textit{Multi-eHMI}, revealed a significant effect of eHMI presence (F(1,18) = 6.253, p = 0.023), with \textit{Multi-eHMI} resulting in lower workload. A Wilcoxon signed-rank test comparing \textit{Multi-eHMI} and \textit{Single-eHMI} also found a significant difference (Z = -2.628, p = 0.009), with \textit{Single-eHMI} further lowering workload.

Qualitative feedback: 
In \textit{Baseline}, participants felt `overwhelmed' (n=6) and `mentally exhausted' (n=12) as they expended excessive effort on `avoiding collisions' and `guessing AV intentions' rather than focusing on navigation. The high mental workload resulted in reduced task efficiency. 
In contrast, in \textit{Multi-eHMI}, PaveFlow's `intuitive' (n=12) signals allowed pedestrians to easily interpret AV intentions, reducing task complexity and enabling greater focus on navigation, significantly improving mental workload and reducing frustration. 
In \textit{Single-eHMI}, the reduced number of AVs further eased task demands, making navigation even smoother.

\subsection{User Experience}

A one-way repeated measures ANOVA comparing \textit{Baseline} and \textit{Multi-eHMI}, revealed a significant effect of eHMI presence (F(1,18) = 15.143, p = 0.001) on pragmatic quality and a significant effect (F(1,18) = 7.674, p = 0.013) on hedonic quality, with \textit{Multi-eHMI} yielding higher scores in both measures. A Wilcoxon signed-rank test comparing \textit{Multi-eHMI} and \textit{Single-eHMI} showed no significant difference on pragmatic quality (Z = 1.454, p = 0.146) and hedonic quality (Z = -0.456, p = 0.649).

Qualitative feedback: In \textit{Baseline}, participants found navigation complicated (n=15). In \textit{Multi-eHMI}, PaveFlow was praised for `providing valuable information,' (n=10) improving pragmatic quality. The dynamic changes in PaveFlow also sparked curiosity, with participants stating `I feel like it reacts because of me,' (P12) emphasising the role of dynamic feedback in enhancing hedonic quality. In \textit{Single-eHMI}, participants described the experience as `most supportive' (n=5) and `easiest,' (n=16) considering it the most user-friendly condition. Interestingly, the perception of right-of-way priority is crucial, as it influences how pedestrians and AVs interact and coexist in shared spaces.
Some participants (n=8) felt that pedestrians should be prioritised in shared spaces traffic rules. They felt PaveFlow’s paths invaded their space, challenging their right-of-way and reducing their satisfaction.
Conversely, participants (n=6) believed that without clear rules in shared spaces, neither pedestrians nor vehicles have absolute priority, suggesting mutual yielding as a courteous gesture.
Yielding to AVs was viewed as a generous act, enhancing their satisfaction.

\section{DISCUSSION}

\subsection{Effectiveness of PaveFlow in AV-Pedestrian Interaction}

\textit{Multi-eHMI} significantly outperformed the \textit{Baseline} across measures, robustly supporting Hypothesis 1 (H1). This confirms that PaveFlow, characterised by its continuity, predictability, and flexible right-of-way, effectively enhances interactions between AVs and pedestrians in shared spaces. 
These findings confirm prior research showing that eHMIs enhance pedestrian safety perception, trust, and decision-making in AV interactions~\cite{hollander2019investigating, faas2020external, rouchitsas2019external}. While eHMIs are generally preferred over baseline conditions without explicit communication, our study extends these insights to shared spaces with multiple AVs, where pedestrian movement is more dynamic and right-of-way is often ambiguous. 

However, qualitative findings indicated that when multiple AVs simultaneously change their motion states—such as accelerating, decelerating, stopping, or adjusting their trajectories—PaveFlow's path projection proved less intuitive and offers limited benefits in reducing cognitive workload. This reveals a key challenge with dynamic feedback: while continuous path updates aid in anticipating AV movements, rapid or simultaneous changes can overwhelm pedestrians, complicating path interpretation.
This observation underscores a critical design challenge for eHMIs in shared spaces with multiple AV—how to maintain clarity and usability during AV motion state transitions~\cite{verma2019pedestrians}.

\subsection{Scalability of PaveFlow in Complex Environments}

\textit{Multi-eHMI} and \textit{Single-eHMI} showed somewhat similar ratings for perceived safety, trust, and user experience, with a slight preference for \textit{Multi-eHMI}. However, notable differences in workload were found, suggesting limitations in PaveFlow's scalability in more complex environments, leading to the rejection of Hypothesis 2 (H2). This was further supported by interview findings, where participants noted that denser path projections and motion state transitions in \textit{Multi-eHMI} complicated decision-making compared to \textit{Single-eHMI}. 

A closer examination of these challenges reveals two paradoxes. First, signals perceived as intuitive in \textit{Single-eHMI} were seen as less intuitive in \textit{Multi-eHMI}. The intuitiveness of eHMI diminishes as the number of eHMIs increases, highlighting the importance of evaluating the scalability of eHMI design concepts~\cite{dey2021towards,tran2023scoping}. Second, the flexible right-of-way principle of the design, intended to enhance AV-pedestrian interactions in complex environments, becomes problematic in \textit{Multi-eHMI}. Real-time eHMI responses, while promoting responsiveness, also increased pedestrian workload~\cite{eisma2023should}.
One potential improvement is adapting the timing of PaveFlow’s projected paths based on the density of AVs and pedestrians. For example, in high-density environments, a slightly delayed or smoothed path projection could prevent excessive, rapid signal updates that may overwhelm pedestrians. 

Finally, to put these challenges of \textit{Multi-eHMI} into perspective and better understand their impact, we need to look at the findings on the effectiveness of \textit{Multi-eHMI} compared to \textit{Baseline} (H1) and \textit{Single-eHMI} (H2) holistically, as they reveal an important nuance of scalability: while the effectiveness of eHMIs diminishes as their number increases, their presence still offers clear benefits over having no eHMI at all. This nuance meaningfully contributes to the ongoing debate over whether eHMIs are merely a gimmick or a necessity~\cite{de2022external}.

\subsection{Study Limitations}
The controlled VR experiment may not fully capture real-world factors such as weather, road surfaces, and unpredictable pedestrian behaviours, which may impact PaveFlow's effectiveness. Technical limitations, such as processing delays and projection clarity in real-world settings, were also not considered. Regarding sample size, the culturally specific participant pool and demographic homogeneity (e.g., strong gender imbalance) may limit generalisability. Finally, while PaveFlow was compared to a no-eHMI baseline, the absence of comparisons with existing eHMI designs makes it unclear whether PaveFlow provides significant advantages over alternative communication methods.

\section{CONCLUSION}

This paper presents the design and evaluation of PaveFlow, a projection-based eHMI that externalises an AV’s intended path and dynamically adapts to pedestrians' movements in shared spaces. Results show that PaveFlow significantly improves AV-pedestrian interactions in multi-vehicle settings compared to scenarios without it. However, as the number of AVs increases, its scalability is challenged by the increased workload caused by motion state transitions, highlighting the need to reassess PaveFlow's communication design and explore solutions to mitigate its negative impacts. This study contributes to a broader understanding of eHMI scalability and the design of effective communication strategies for AVs in shared spaces.

\begin{acks}
We sincerely thank the capstone research unit coordinator Joel Fredericks for his guidance and support, the participants for their valuable contributions, and the anonymous reviewers for their insightful feedback. This research is supported by the Australian Research Council (ARC) Discovery Project DP220102019, Shared-space interactions between people and autonomous vehicles.
\end{acks}

\bibliographystyle{ACM-Reference-Format}
\bibliography{references}


\end{document}